\documentclass{iau} 
\usepackage{graphicx}

\title[The enigmatic winds of Wolf-Rayet stars] 
{The enigmatic winds of Wolf-Rayet stars: Results from dynamically consistent atmosphere modelling}

\author[A.A.C. Sander]   
{Andreas A.C. Sander$^1$}

\affiliation{$^1$Zentrum für Astronomie der Universität Heidelberg, Astronomisches Rechen-Institut, Mönchhofstr. 12-14, 69120 Heidelberg, Germany \\ email: {\tt andreas.sander@uni-heidelberg.de} 
}

\pubyear{2022}
\volume{361}  
\jname{Massive Stars Near and Far}
\editors{N.\ St-Louis, J.\ S.\ Vink, J.\ Mackey, eds.}
\begin{document}

\maketitle

\begin{abstract}
Line-driven stellar winds are ubiquitous among hot massive stars. In some cases they can become so strong, that the whole star is cloaked by an optically thick wind. The strong outflow gives rise to large emission lines, defining the class of so-called Wolf-Rayet (WR) stars. While being major players in the evolution of massive stars, the formation of heavy black holes, and the distribution of elements, the occurrence and nature of WR winds is still quite enigmatic.

A promising instrument towards a better theoretical understanding are stellar atmospheres allowing for a consistent inclusion of the hydrodynamics. By coupling stellar and wind parameters and the inclusion of a detailed non-LTE radiative transfer, they allow us to go beneath the observable layers and study the onset of WR-type winds. Establishing larger sets of models, we were able to make ground-breaking progress by identifying trends with mass and metallicity that deviate significantly from present empirical descriptions.
Our modelling efforts reveal a complex picture for WR-type winds with strong, non-linear dependencies. Besides covering metallicity and mass, we further identify surface hydrogen as an important ingredient to retain WR-type mass loss at lower metallicity. Here, we present a summary of recent insights on the nature and onset of WR-type winds in massive stars including the consequences for stellar evolution, remaining open questions, and current efforts to overcome them.

\keywords{stars: atmospheres, stars: mass loss, stars: massive, stars: winds, outflows, stars: Wolf-Rayet, stars: evolution, stars: black holes, galaxies: stellar content}
\end{abstract}

\firstsection 
\section{Introduction}

The Wolf-Rayet (WR) phenomenon indicates the presence of strong winds that can partially or completely hide the underlying stellar surface. Consequently, the connection of stellar and wind parameters has been a challenge for these objects due to inherent degeneracies in the spectrum formation \cite[(e.g., Hamann \& Gr{\"a}fener 2004; Lefever et al., these proceedings)]{HamannGraefener2004,Lefever+2022}. Motivated by theoretical considerations of optically thick winds launched in the deeper atmospheres \cite[(e.g., Nugis \& Lamers 2002)]{NugisLamers2002}, the introduction of hydrodynamically consistent stellar atmosphere models for WR stars \cite[(Gr{\"a}fener \& Hamann 2005, Sander et al. 2020)]{GraefenerHamann2005,Sander+2020} has opened a new window to overcome these issues. Already the first model by \cite[Gr{\"a}fener \& Hamann (2005)]{GraefenerHamann2005} reprising a ``classical'', i.e.\ helium-burning, WC star confirmed the suggestion of \cite[Nugis \& Lamers (2002)]{NugisLamers2002} that the opacities of the iron M-shell ions are fundamental to launch and drive the winds of classical WR stars. At the same time, independent Monte Carlo simulations at a lower temperature regime by \cite[Vink \& de Koter (2005)]{VinkdeKoter2005} also yielded a fundamental Fe-dependence for WR-type winds.

\begin{figure}[htb]
\begin{center}
   \includegraphics[width=0.40\textwidth]{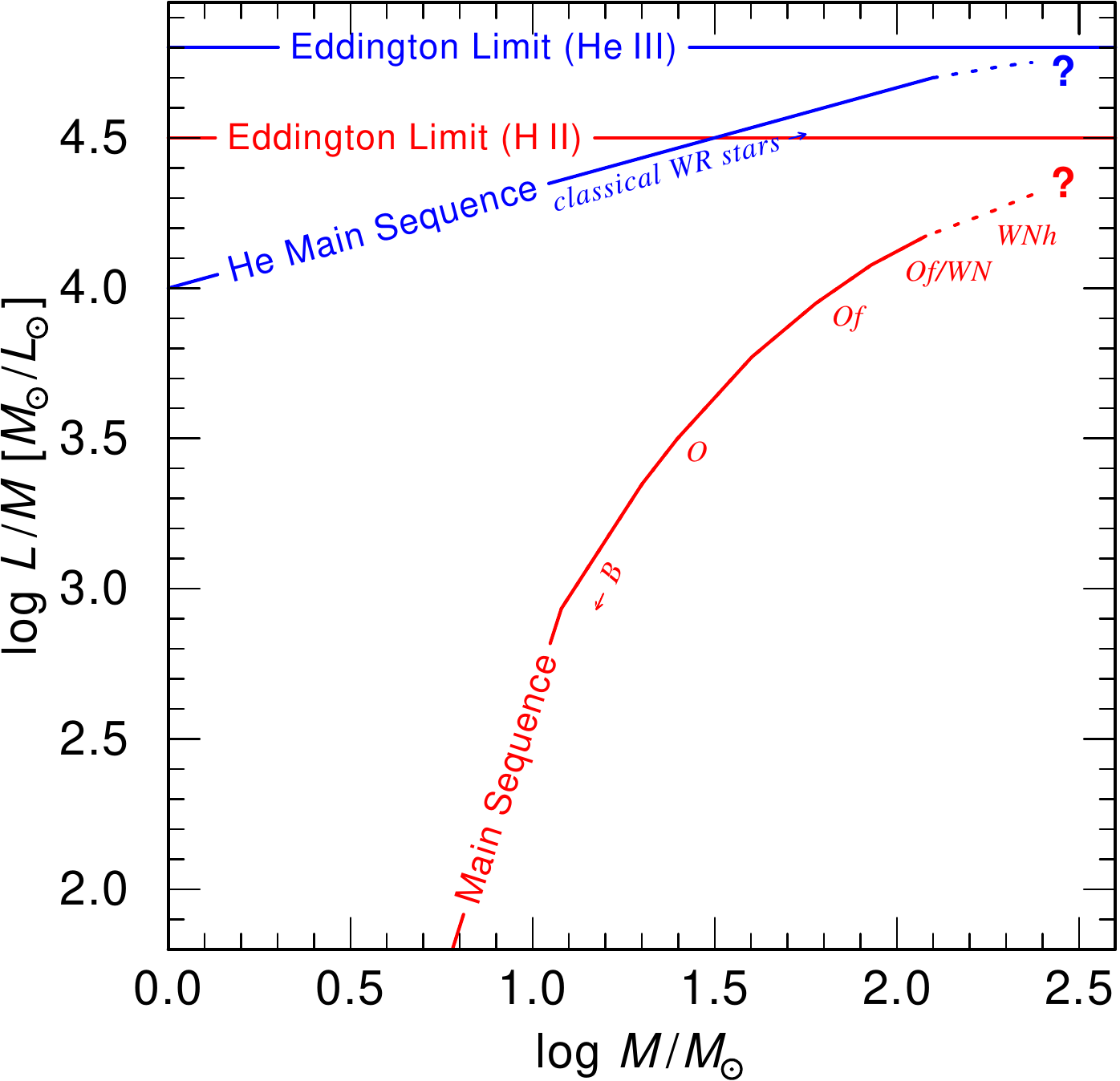} \hfill
 	 \includegraphics[width=0.57\textwidth]{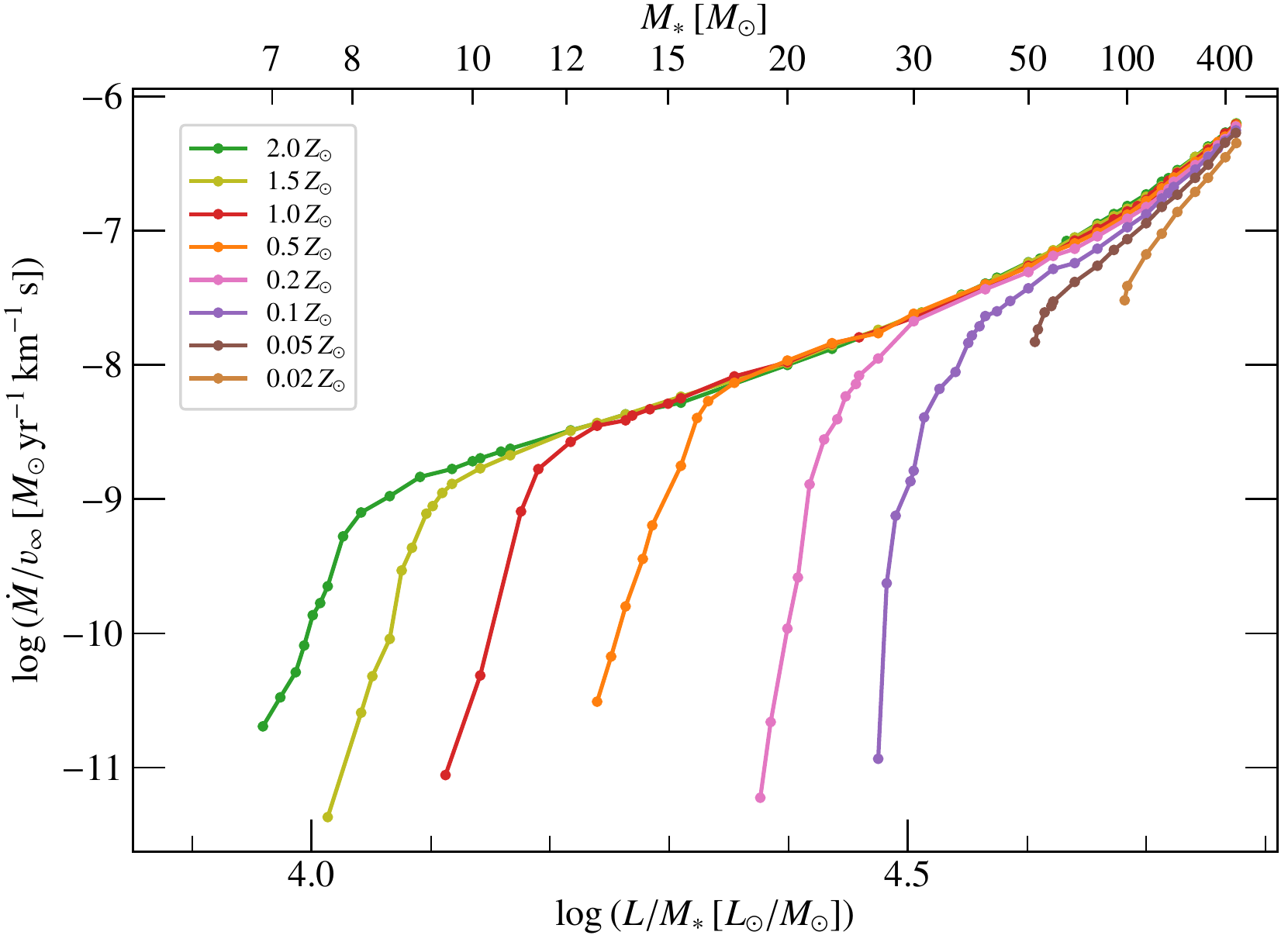}
 \caption{Left panel: Schematic occurrence of WR-type spectra in the $\log L/M$ vs.\ $\log M$ plane. When approaching the Eddington limit for hydrogen-containing stars at the upper end of the main sequence, we see a transition from Of to WN spectral type. Classical WR stars located on the He main sequence are close to the Eddington limit for hydrogen-free stars and thus also show strong stellar winds.
  Right panel: Ratio of the mass-loss rate $\dot{M}$ and the terminal velocity $v_\infty$ as a function of $L/M$ for the models from \cite[Sander \& Vink (2020)]{SanderVink2020}.}
   \label{fig:wrtype-mdot}
\end{center}
\end{figure}

For the cooler, very massive WNh stars, \cite[Gr{\"a}fener \& Hamann (2008)]{GraefenerHamann2008} calculated grids of models.
Together with the work of \cite[Vink et al. (2011)]{Vink+2011}, their study demonstrated that the increased mass loss of WR stars is a natural consequence when stars approach the Eddington limit of $\Gamma_\mathrm{e} \rightarrow 1$.
The corresponding limit in $L/M$ differs between hydrogen-free and hydrogen-rich stars due to $\Gamma_\mathrm{e} \propto q_\mathrm{ion} L/M$ \cite[(see, e.g., Sander et al. 2015 for detailed formulae)]{Sander+2015} as hydrogen changes the value of $q_\mathrm{ion}$. This is schematically illustrated in the left panel of Fig.\,\ref{fig:wrtype-mdot}, where we see the trends in $L/M$ versus $M$ for the main sequence and the He main sequence including their corresponding Eddington limits in $L/M$. 

\section{Mass-loss rates for classical WN stars}

For the hotter, classical WR stars on the He main sequence, \cite[Sander \& Vink (2020)]{SanderVink2020} calculated the first large set of hydrodynamically-consistent models. Employing the mass-luminosity relation from \cite[Gr{\"a}fener et al.\ (2011)]{Graefener+2011} to reduce the parameter space and fixing the inner boundary constrain $T_\ast(\tau_\mathrm{Ross} = 20) = 141\,$kK, they provide a description of $\dot{M}$ from sequences in $L/M$ at multiple metallicities $Z$.
A summary of the \cite[Sander \& Vink (2020)]{SanderVink2020} model results is depicted in the right panel of Fig.\,\ref{fig:wrtype-mdot}, where the ratio of $\dot{M}$ and the terminal wind velocity $v_\infty$ is shown as a function of $L/M$. Surprisingly and very much in contrast to OB-type winds, the curves for different metallicities start to align in the limit of dense winds, which is not the case when looking only at $\dot{M}$. The slope of these aligned curves can be brought into a linear dependence when plotting against $\log\left[-\log\left(1-\Gamma_\mathrm{e}\right)\right]$. Another important result from the models is a $Z$-dependent breakdown of $\dot{M}$ when the sonic point of the wind approaches the optically thin regime. At higher $Z$, strong mass-loss can be maintained down to lower $L/M$ ratios as more line opacities are available. The finding of this rapid, $Z$-dependent breakdown qualitatively explains the empirical lower cutoffs found in observed WN populations of different $Z$ \cite[(cf.\ Shenar et al.\ 2020)]{Shenar+2020}. 

As illustrated later on in the WN-WC comparison (cf.\ right panel of Fig.\,\ref{fig:wc-wn-cmp}), the breakdown of $\dot{M}$ is accompanied by a change in the trend for $v_\infty$, which reaches a minimum before eventually taking a steep turn towards higher values once the winds become optically thin. This result is tied to the choice of $T_\ast$ (and thus the flux at the inner boundary). With our high value of $141\,$kK, the obtained thin-wind solutions are still launched by the opacities of the high Fe bump \cite[(as illustrated in Sander et al.\ 2020)]{Sander+2020}. The conditions in the resulting hot and thin winds are sufficient to keep Fe (and other elements) at higher ionization stages for thousands of stellar radii. In such cases, the acceleration of the thin, fast wind can be maintained and there is no lower limit in $\dot{M}$, in contrast to the results obtained in stellar structure calculations \cite[(e.g.\ Grassitelli et al.\ 2018, Ro 2019)]{Grassitelli+2018,Ro2019}. It is presently unclear whether such massive, luminous, and hot stars do exist in nature or if their formation is prohibited by structural constraints. The detection of these objects, especially in unresolved low-$Z$ galaxies, would be non-trivial as their predicted spectra neither show strong optical emission lines, nor pronounced P\,Cygni profiles in the UV.

\section{Influence of leftover hydrogen at the surface}
  \label{sec:hydrogen}

\begin{figure}[htb]
\begin{center}
   \includegraphics[angle=270,width=\textwidth]{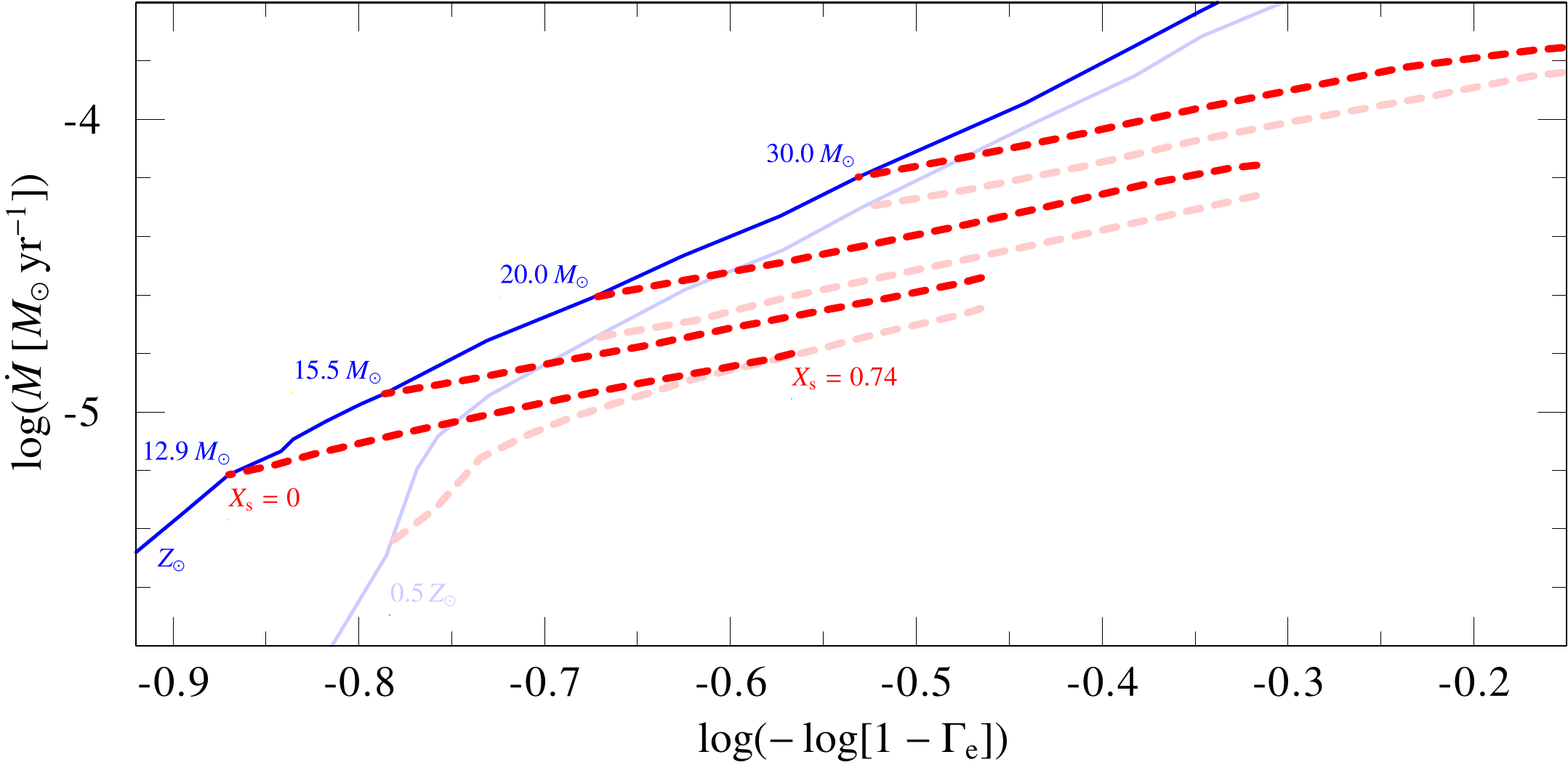}
     \caption{Influence of remaining surface hydrogen on the mass-loss rate $\dot{M}$: The blue curves show our sequence of hydrogen-free WN-type models with increasing $L/M$. For selected ($L$,$M$)-sets, the trend of including hydrogen in the atmospheric calculations (at the cost of helium) is illustrated by red, dashed curves. The opaque lines denote models at $Z_\odot$, while the semi-transparent lines refer to the same set of calculations as $0.5\,Z_\odot$.}
   \label{fig:surface-h}
\end{center}
\end{figure}

In addition to the two-dimensional parameter study along $L/M$ and $Z$ published in \cite[Sander \& Vink (2020)]{SanderVink2020}, we present first results from our work accounting for remaining hydrogen on the surface of WN-type stars. For the results depicted in Fig.\,\ref{fig:surface-h}, the models assume that the total hydrogen content is negligible in regard to the stellar mass, meaning that $L$ and $M$ are identical to the hydrogen-free calculations. In nature this would correspond to stars having only a thin, hydrogen-containing layer on the surface.

We compare the general trend of $\dot{M}$ for increasing $L/M$ with the trends arising when increasing the surface hydrogen mass fraction $X_\mathrm{H}$ for two different metallicities applying models with fixed $L$, $T_\ast$, and $M$ in Fig.\,\ref{fig:surface-h}. In the limit of optically thick winds, we always obtain the linear scaling of $\log \dot{M}$ with $\log\left[-\left(1-\Gamma_\mathrm{e}\right)\right]$ found in \cite[Sander \& Vink (2020)]{SanderVink2020}. Interestingly, the slope of the curves where we vary $X_\mathrm{H}$ is significantly shallower than the curve denoting the trend with varying $L/M$. The $20\,M_\odot$ model at $Z_\odot$ with $X_\mathrm{H} = 0.2$ has approximately the same $\Gamma_\mathrm{e}$ as the $30\,M_\odot$ model with $X_\mathrm{H} = 0$, but its mass-loss rate is a factor of two lower. The terminal velocity of the hydrogen-containing model is also about $250\,\mathrm{km}\,\mathrm{s}^{-1}$ lower than the hydrogen-free $30\,M_\odot$ model, which is contrary to what happens when carbon is increased at the cost of helium as we will discuss further below.   

Independent of our calculations representing classical WR stars, \cite[Sabhahit et al.\ (2022)]{Sabhahit+2022} recently found qualitatively similar results from stellar evolution calculations for very massive (WNh-type) stars. In their calculations, they needed the scaling of $\dot{M}$ with $X_\mathrm{H}$ to be weaker than the scaling with $L/M$ in order to ensure that the resulting evolutionary tracks were covering only the observed region of WNh stars in the HRD. Otherwise, the mass loss would be too weak to prevent stars from expanding to cooler, unobserved temperatures. We are presently following up these findings with dynamically consistent calculations in the very massive star regime in order to check whether our insights on $\dot{M}(X_\mathrm{H})$ for classical WR are also applicable in this different temperature and mass regime or if the scaling diverges otherwise.

\section{Similarities between WN and WC stars}
  \label{sec:wnwc}

As discussed above, the winds of WR stars are mainly launched by Fe M-shell opacities in the calculated temperature regime of $T_\ast \approx 140\,$kK. In the outer winds, Fe ions usually remain the leading opacity for further accelerating the wind, but are supported by a plethora of further elements including carbon and oxygen. An immediate consequence of this situation is a strong similarity in the $\dot{M}$-dependences of WN and WC stars.

\begin{figure}[htb]
\begin{center}
   \includegraphics[width=0.478\textwidth]{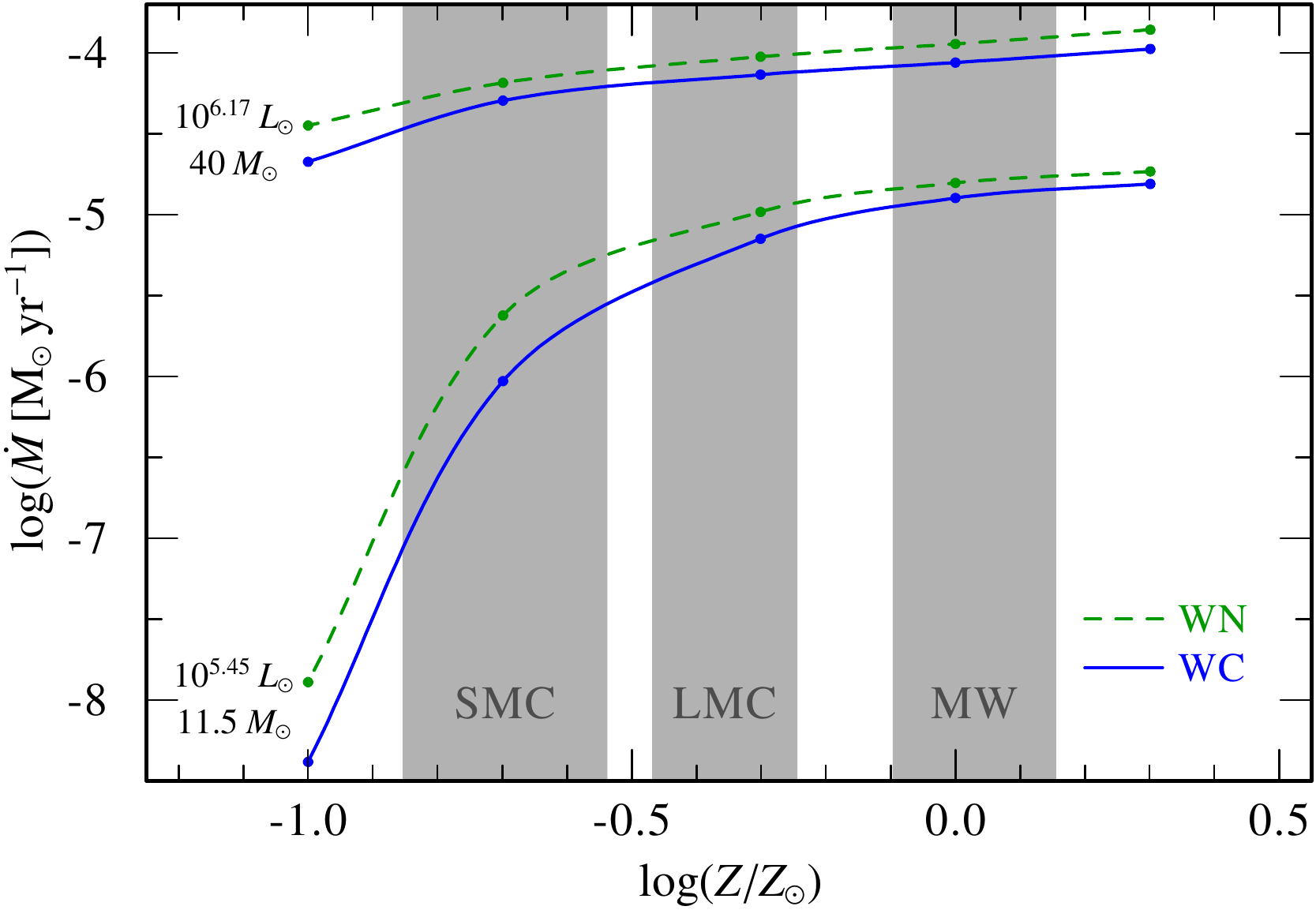} \hfill
   \includegraphics[width=0.50\textwidth]{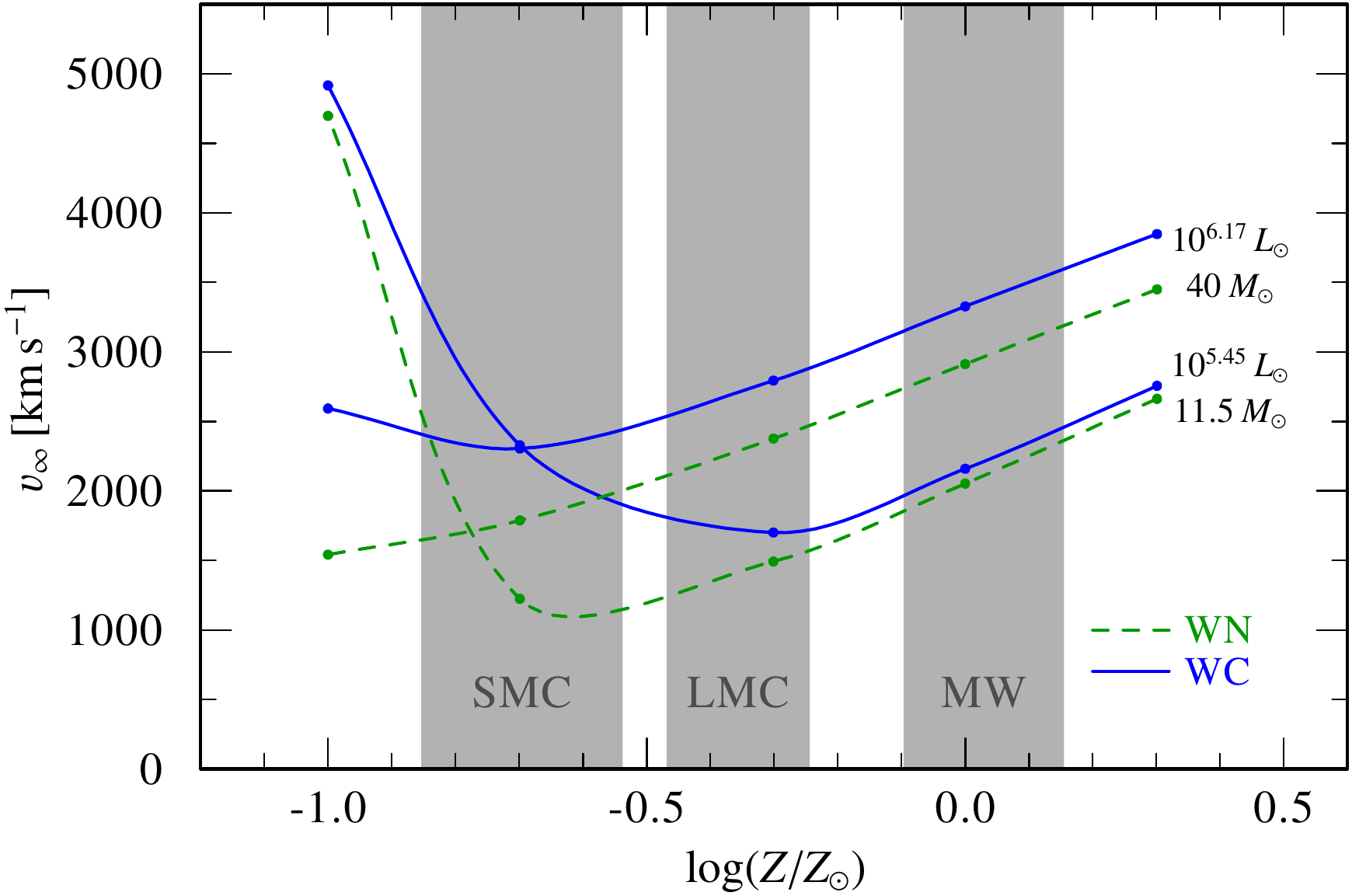} 
     \caption{Mass loss rates (left panel) and terminal velocities (right panel) as a function of metallicity $Z$ for four different model sequences. For each of the two different sets of basic stellar parameters ($L$, $M$), two different sets of surface abundances representing WN (green, dashed) and WC (blue, solid) stars were calculated.}
   \label{fig:wc-wn-cmp}
\end{center}
\end{figure}

In Fig.\,\ref{fig:wc-wn-cmp}, we illustrate this similarity by showing WN and WC model sequences for two different sets of basic stellar parameters (both assuming $T_\ast = 141\,$kK). For the mass-loss rate, the WC model curves closely resemble the WN model curves, but with slightly \textit{lower} absolute values for $\dot{M}$. The latter difference mainly roots in the slightly lower amount of free electron scattering available (and thus lower $\Gamma_\mathrm{e}$) in the WC models. In the WN models, helium has a fraction of $0.95$ to $0.99$ of the atmospheric mass (with detailed values depending on the metallicity of the model). These values are reduced by $0.45$ ($40\,M_\odot$ sequence) and $0.65$ ($11.5\,M_\odot$ sequence) in the WC models where we assume $X_\mathrm{C} = 0.4$ or $0.6$ and $X_\mathrm{O} = 0.05$ for carbon and oxygen respectively. Although both C and O are in higher ionization stages at the launch of the wind and thus do not provide considerable line-driving opacities, both of these elements are not fully ionized. Hence, the budget of free electrons is reduced in comparison to the WN models and consequently there is less total acceleration available when adding up line and continuum opacities in the deeper layers. This leads to a slightly lower $\dot{M}$ for the same $L$ and $M_\ast$, in contrast to often applied empirical recipes such as \cite[Nugis \& Lamers (2000)]{NugisLamers2000}. However, when comparing WN and WC models with the same $\Gamma_\mathrm{e}$, the WC model yields the higher $\dot{M}$ (cf.\ Fig.\ 15 in \cite[Sander et al.\ 2020]{Sander+2020}).

A second effect of changing the surface abundances from a WN to a WC composition is shown in the right panel of Fig.\,\ref{fig:wc-wn-cmp}: As the higher $X_\mathrm{C}$ and $X_\mathrm{O}$ provide more line opacities in the wind, the WC models show a higher $v_\infty$ than the WN models. This difference gets stronger for higher $L/M$, i.e.\ when getting closer to the Eddington limit. The examples further illustrate that the minimum of the $v_\infty$-curves is different for all four depicted sequences. As this minimum is related to the transition to optically thin winds, it underlines that this transition is not only affected by changing basic parameters such as $L$ and $M$, but also by (significant) abundance changes. Due to the generally lower $\dot{M}$ in the WC models, the transition to higher $v_\infty$ occurs already at higher metallicities.

\section{Summary \& Conclusions}

By calculating sequences of hydro-dynamically consistent atmosphere models along multiple dimensions, we obtained the first locally consistent $\dot{M}$-description for classical WR stars based only on first principles. The models further reveal that WR-type winds differ in various aspects substantially from OB-type winds. With increasing wind strength, we reach an optically thick (``pure WR'') regime where the sonic point of WR-type wind located very deep in the atmosphere where the conditions do no longer deviate significantly from LTE \cite[(Sander \& Vink 2020)]{SanderVink2020}. However, the outer wind regime, which determines $v_\infty$, remains in strong non-LTE, preventing any straight-forward simplification of the wind calculations.
The precise dependencies of $\dot{M}$ as a function of stellar parameters are complex, although clear patterns emerge from calculating series of models along the $L/M$- and $Z$-dimension. In the limit of the pure WR regime, we find $\dot{M} \propto 1/\log(1-\Gamma_\mathrm{e})$ and a surprisingly shallow metallicity-dependence of $\dot{M} \propto Z^{0.3}$. In \cite[Higgins et al. (2021)]{Higgins+2021} we performed a first test of implementing the new $\dot{M}$-description into stellar evolution models of He stars, generally yielding lower total mass loss and potentially allowing for pair-instability supernovae up to $0.5\,Z_\odot$. Significant open questions remain, including the importance of further dependencies (e.g.\ $T_\ast$, $X_\mathrm{H}$), where our calculations hint that additional terms need to be incorporated into future $\dot{M}$-descriptions. Eventually, a coherent picture also has to take into account the multi-dimensional WR wind structure, for which intriguing new insights are coming from recent 3D calculations \cite[(e.g.\ Moens et al.\ 2022)]{Moens+2022}. Together with colleagues, we will test and extend current approximation techniques \cite[(e.g. Poniatowski et al.\ 2021)]{Poniatowski+2021} to bring 1D and 3D modelling closer together and go beyond LTE in 3D calculations as well as implement 3D-insights into 1D atmosphere modelling.

\begin{discussion}

\discuss{Stevance}{I have a question about your statement that some surface hydrogen might increase the mass loss: What do you mean by a negligible mass fraction?}

\discuss{Sander}{By negligible I mean that it does not change the $L/M$-ratio when you compare it to a hydrogen-free star. This is what we have done in the presented models. Of course, if you put a considerable amount of hydrogen-rich mass on top of a helium star, the effect of the higher gravity will take over and your mass-loss rate should go down. We will also work on models with non-negligible mass additions, but this adds a further dimension and thus will take time. Generally you can expect that for a significant additional mass your mass-loss rate to be lower than the result coming out of the helium star model.}

\discuss{Laplace}{I'm also very excited about the remaining hydrogen. I was wondering that if you have this increased mass loss in case of surface hydrogen, does it mean you will get rid of the hydrogen? This would really matter for the further evolution and end stage of the star including its spectra.}

\discuss{Sander}{As long as the additional hydrogen-rich mass is significant, I would expect that the resulting mass loss is decreased. But indeed the last bit of hydrogen should be very easy to remove because of this enhanced mass-loss effect.}

\discuss{Langer}{Great work, very much needed. With respect to your talk, I also liked that you highlight the importance of the Eddington limit. From the stellar structure considerations, if stars go there you find inflation and I was surprised that this did not come up in your talk. In the old days we had problems to find consistent WR mass-loss rates together with small radii. Are your new helium star models inflated or are they compact?}

\discuss{Sander}{Thank you very much for this excellent question that allows me to cover stuff that I could otherwise not fit in a 12-minute talk. The models that I talked about all have winds launched by the hot iron bump and their sonic points are on the order of $1\,R_\ast$. So they are compact, but they do not appear compact because they have this optically thick inner wind region. The winds will only get optically thin at much larger radii. Thus, from a ``naive'' $\beta$-law analysis, one would conclude that they have inflated radii. What we need to figure out now, also by comparing with observed spectra, is whether the solution of the current models is always the correct one or whether an inflated (hydrostatic) radius is an actual alternative for at least some observed stars. Moreover, there can also be deceleration regions in the wind. We did not see them in the models presented here, but we are now working on further models where we need to do some mitigations in order to treat them in the code. So, I would say the question is not settled yet, but my current feeling is that the winds are actually launched deep, but then sometimes meet a deceleration region. This would mean that we have a moving atmosphere rather than a hydrostatic inflation, but this is just my surmise at the moment.}

\end{discussion}

\end{document}